\documentclass[10pt,twocolumn,superscriptaddress]{revtex4-2}

\usepackage{amsmath,amsfonts}
\usepackage[colorlinks=true,linkcolor=cyan,citecolor=cyan,urlcolor=cyan]{hyperref}

\begin{document}

\title{Estimation of a parameter encoded in the modal structure of a light beam:\\ a quantum theory}

\author{Manuel Gessner}
\affiliation{Departament de F\'isica Te\`orica, IFIC, Universitat de Val\`encia, CSIC, C/ Dr. Moliner 50, 46100 Burjassot (Val\`encia), Spain}
\email{manuel.gessner@uv.es}
\author{Nicolas Treps}
\affiliation{Laboratoire Kastler Brossel, Sorbonne Universit\'e, ENS-Universit\'e PSL, CNRS, Coll\`ege de France, 4 Place Jussieu, F-75252 Paris, France}
\author{Claude Fabre}
\affiliation{Laboratoire Kastler Brossel, Sorbonne Universit\'e, ENS-Universit\'e PSL, CNRS, Coll\`ege de France, 4 Place Jussieu, F-75252 Paris, France}

\begin{abstract}
Quantum light is described not only by a quantum state but also by the shape of the electromagnetic modes on which the state is defined. Optical precision measurements often estimate a ``mode parameter'' that determines properties such as frequency, temporal shape and the spatial distribution of the light field. By deriving quantum precision limits, we establish the fundamental bounds for mode parameter estimation. Our results reveal explicit mode-design recipes that enable the estimation of any mode parameter with quantum enhanced precision. Our approach provides practical methods for optimizing mode parameter estimation with relevant applications, including spatial and temporal positioning, spectroscopy, phase estimation, and superresolution imaging.
\end{abstract}


\maketitle

\section{Introduction}
As a particularly sensitive carrier of information, light represents an exceptional platform for precision measurements, with applications including spectroscopy, interferometry, positioning, timing, and imaging. The properties of quantum light are determined, on the one hand, by its quantum state, which may be coherent, thermal or a Fock state, for instance. On the other hand, a full description of the electromagnetic field further requires knowledge of the modes on which this quantum state is defined~\cite{FabreRMP2020}. These modes are normalized solutions of Maxwell's equations that determine the spatial intensity distribution, frequency spectrum and polarization of the light field. We refer to the parameters that determine these modal properties as mode parameter. In contrast, the number of photons, purity or temperature are parameters that define the state on these modes. Other parameters, such as phase shifts, can be equivalently considered as parameter of the mode or the state.

The ultimate precision limit on the measurement of any parameter can be determined within the framework of quantum metrology~\cite{HelstromBOOK,HolevoBOOK,CavesPRD1981,WinelandPRA1992,ParisINTJQI2009,GiovannettiNATPHOT2011}. A lower bound on the variance of any unbiased estimator for the parameter of interest is given by the quantum Cram\'er-Rao bound. This bound can be achieved asymptotically when a large number of measurement results is available, for instance by applying maximum likelihood estimation to the data obtained from an optimal observable. This approach has led to the improvement of the error scaling in several measurements~\cite{RafalPROGOPT2015,PezzeRMP2018,TsangREVIEW2020}, e.g., in gravitational wave detectors~\cite{TsePRL2019} or atomic clocks and interferometers~\cite{PezzeRMP2018}.

Methods to construct the optimal observable and to determine the quantum Cram\'er-Rao bound are, in principle, available for arbitrary states and parameters. Explicit expressions can be obtained, e.g., from the spectral decomposition of the state~\cite{BraunsteinPRL1994,ParisINTJQI2009,GiovannettiNATPHOT2011,PezzeRMP2018}, from integral representations~\cite{ParisINTJQI2009}, or using matrix vectorization techniques~\cite{SafranekPRA2018}. Furthermore, convenient decompositions in terms of the covariance matrix are available for Gaussian states~\cite{GaussianQFI}. However, all of these techniques require the explicit description of the evolution of the quantum state under variations of the parameter. This is particularly simple when the parameter is imprinted by a unitary transformation and the evolution is given in terms of a parameter-independent Hamiltonian via Schr\"odinger's equation. But when the parameter of interest describes a property of the mode, the resulting evolution of the state is less evident.

So far, mode parameter estimation has been approached on a case-by-case basis. Known results that explicitly discuss the precision limits for a mode parameter are either limited to specific observables and estimators~\cite{FabreOPTLETT2000,TrepsPRA2005,LaminePRL2008}, or to specific states, such as a combination of a coherent state with a pure Gaussian state~\cite{PinelPRA2012}. A notable recent breakthrough was the identification of sub-Rayleigh imaging strategies that outperform traditional direct imaging~\cite{TsangPRX2016,TsangREVIEW2020} through a systematic optimization over all measurement observables. Formally, the imaging problem addresses the estimation of the separation of the two source's transverse spatial modes. The quantum limit was first derived in the limit of very faint sources~\cite{TsangPRX2016}, and has since been generalized to arbitrary number-diagonal states~\cite{LupoPRL2016}, which include thermal states~\cite{TsangPRL2016}.

One of the most interesting prospects of quantum metrology is the improvement of the error scaling beyond the standard quantum limit (SQL). Quantum-enhanced measurement strategies exploit squeezed or entangled states~\cite{CavesPRD1981,WinelandPRA1992,PezzePRL2009,RivasPRL2010,PezzeRMP2018} in order to reduce the relevant quantum fluctuations beyond those of the vacuum, which define the SQL. Several results in the existing literature suggest that such enhancements are impossible for mode parameter estimation problems: Known precision bounds, e.g., in superresolution imaging~\cite{LupoPRL2016} and the estimation of a beam displacement with a single mode field~\cite{FabreOPTLETT2000}, depend only on the average number of photons $N=\langle \hat{N}\rangle$. Quantum strategies can suppress measurement fluctuations beyond the SQL only if the precision bound actually depends on these fluctuations, e.g., when terms such as $\langle \hat{N}^2\rangle$ are present. In such cases, which include interferometric measurements, quantum metrology offers a wealth of well-known strategies ranging from squeezed to NOON states that lead to improved precision~\cite{GiovannettiNATPHOT2011,PezzeRMP2018}. The maximal theoretically possible improvement is known as the Heisenberg limit and corresponds to a factor of $N$ over the SQL. For the estimation of a beam displacement, it was shown that by populating a second, carefully designed mode, quantum enhancements are possible~\cite{FabreOPTLETT2000,TrepsSCIENCE2003}. 



In this article, we derive a quantum theory that identifies the quantum precision limit on the estimation of any mode parameter by optimizing over all possible quantum measurements without making assumptions about the state or the modes on which it is defined. We find that any mode parameter can be estimated with quantum-enhanced precision if suitable modes are populated with nonclassical states. Our results reveal mode design strategies that enable a scaling improvement which consist in either (1) choosing initial modes that have nonvanishing overlap with their own derivative modes or (2) populating suitably chosen auxiliary modes. Both approaches open the way to reduce measurement noise below the SQL using standard techniques based on squeezed or other nonclassical states. We illustrate the general applicability of our framework with various examples, including the estimation of beam displacements and superresolution imaging.

\textbf{Quantum theory of mode parameter estimation.} The quantum limit on the estimation of a parameter $\theta$ is given by the quantum Cram\'er-Rao bound $(\Delta \theta_{\mathrm{est}})^2\geq 1/F_Q[\hat{\rho}(\theta)]$, where the quantum Fisher information (QFI) $F_Q[\hat{\rho}(\theta)]$~\cite{HelstromBOOK,BraunsteinPRL1994,ParisINTJQI2009,PezzeRMP2018} describes the sensitivity of the state $\hat{\rho}(\theta)$ under variations of $\theta$. The central quantity of interest, the QFI, can be determined explicitly with a variety of methods~\cite{BraunsteinPRL1994,ParisINTJQI2009,GiovannettiNATPHOT2011,SafranekPRA2018,GaussianQFI}, which all require explicit knowledge of $\frac{\partial}{\partial \theta}\hat{\rho}(\theta)$. The evolution $\frac{\partial}{\partial \theta}\hat{\rho}(\theta)$ describes the variation of the quantum state under changes of the parameter of interest. Note that it is generally not necessary to vary the parameter $\theta$ in an experiment. Typically, we detect small deviations of $\theta$ around a fixed value that we can use to define $\theta=0$.

A particularly simple situation arises when the parameter is imprinted via a unitary transformation with a parameter-independent Hamiltonian $H$. In this case, $\frac{\partial}{\partial \theta}\hat{\rho}(\theta)=-i[\hat{H},\hat{\rho}(\theta)]$ is governed by the von Neumann equation and explicit expressions for the QFI $F_Q[\hat{\rho},\hat{H}]$ are available as a function of the initial state $\hat{\rho}=\hat{\rho}(0)$ and the Hamiltonian $\hat{H}$. The QFI $F_Q[\hat{\rho},\hat{H}]$ expresses the quantum fluctuations of the state $\hat{\rho}$ and, indeed, depends quadratically on $\hat{H}$. In fact, for pure states $\hat{\psi}$, the QFI for unitary evolutions reduces to the variance: $F_Q[\hat{\psi},\hat{H}]=4(\Delta\hat{H})^2_{\hat{\psi}}=4(\langle\hat{H}^2\rangle_{\hat{\psi}}-\langle\hat{H}\rangle_{\hat{\psi}}^2)$~\cite{BraunsteinPRL1994}. The nonlinear term $\langle\hat{H}^2\rangle_{\hat{\psi}}$ can be exploited to reduce the quantum noise beyond the SQL with the help of nonclassical states, such as metrologically useful entangled states or squeezed states whose quantum fluctuations are smaller than those of coherent and vacuum states~\cite{CavesPRD1981,WinelandPRA1992,PezzePRL2009,RivasPRL2010,PezzeRMP2018}. 

Let us now address the variation of a quantum state $\frac{\partial}{\partial \theta}\hat{\rho}(\theta)$ that is caused by changes of a parameter $\theta$ that determines the modes on which the state is defined. We focus on the common situation where a parameter-independent quantum state is prepared in one or several modes whose shape depends on a single parameter of interest. Consider a mode basis $\{f_k\}_{k}$ with an inner product $(f_k|f_l)=\int dx f_{k}^*(x)f_l(x)=\delta_{kl}$. Here, $x$ denotes an abstract set of arguments of the modes, which may be, e.g., spatial, temporal, frequency or other coordinates. A perturbation of these modes gives rise to the shifted mode basis $\{f_{k}[\theta]\}_k$, parametrized by $\theta$, such that at $\theta=0$ we recover the original modes, i.e., $f_{k}[0]=f_k$ for all $k$. 

As our first result, we show that changes of an arbitrary state $\hat{\rho}$ due to a variation of the mode parameter $\theta$ around $\theta=0$ can be described by a unitary beam-splitter evolution (see Supplemental document) as $\frac{\partial}{\partial \theta}\hat{\rho}=-i[\hat{H},\hat{\rho}]$, with the effective Hamiltonian
\begin{align}\label{eq:H}
    \hat{H}&=i\sum_{jk}(f_j|f'_{k})\hat{a}_j^{\dagger}\hat{a}_k = \sum_{k}w_k\hat{d}_k^{\dagger}\hat{a}_k.
\end{align}
Here, $\hat{a}_k^{\dagger}$ and $d^{\dagger}_k=\frac{i}{w_k}\sum_{j}(f_j|f'_{k})\hat{a}_j^{\dagger}$ create a photon in the mode $f_k$ and $\frac{if'_{k}}{w_k}=\frac{i}{w_k}\frac{\partial}{\partial \theta}f_{k}[\theta]|_{\theta=0}$, respectively, where $w_k=\sqrt{(f'_k|f'_k)}$. The resulting mode-mixing evolution thus coherently redistributes populations from the original modes $f_k$ into the derivative modes $\frac{if_k'}{w_k}$. The coupling coefficients of the effective Hamiltonian depend on the shape and normalization $w_k$ of the modes via the overlap integral $i(f_j|f'_k)$. Note that orthonormality of the modes implies that $(f_k'|f_j)=-(f_k|f'_j)$ and thus $\hat{H}$ is Hermitian.

The effective Hamiltonian~(\ref{eq:H}) provides a remarkably simple description of a quantum state's dependence on a mode parameter and translates an evolution of the modes into an evolution of the state. Most importantly in our context, this result allows us to determine the quantum limits on any mode parameter estimation using well-known expressions for the QFI $F_Q[\hat{\rho},\hat{H}]$ for unitary evolutions, which apply to arbitrary quantum states.

In practical situations, typically only a finite number of the modes $f_k$ will be occupied by the initial state $\hat{\rho}$. By explicitly distinguishing populated modes from vacuum modes, we gain insight into the strategies that allow us to optimize the quantum limits on mode parameter estimation. To this end, let us introduce
\begin{align}\label{eq:HI}
    \hat{H}_I=i\sum_{jk\in I}(f_j|f'_k)\hat{a}_j^{\dagger}\hat{a}_k,
\end{align}
where $I$ is the set of modes that is occupied by the state $\hat{\rho}$. We obtain our main result (see Supplemental document):
\begin{align}\label{eq:QFI}
    F_Q[\hat{\rho}(\theta)]|_{\theta=0}&=F_Q[\hat{\rho},\hat{H}_I]\\&\quad+4\sum_{kl\in I}\left[(f'_k|f'_{l})-\sum_{j\in I}(f'_k|f_j)(f_j|f'_{l})\right]\langle \hat{a}_k^{\dagger}\hat{a}_l\rangle_{\hat{\rho}}.\notag
\end{align}
In this expression, the sensitivity of the state is described exclusively in terms of modes that are initially populated. The first term in~(\ref{eq:QFI}) is the QFI for a unitary evolution generated by $\hat{H}_I$ and thus contains relevant quantum fluctuations of the state $\rho$. As we will detail below, nonclassical states with quantum fluctuations below the SQL are able to generate quantum enhancements by increasing this term beyond classical limits. In contrast, the second term always scales linearly with the average number of photons. It is therefore independent of the state's quantum fluctuations and consequently cannot beat the scaling of the SQL. This less favorable scaling ensues from the quantum noise of vacuum modes that exchange relevant information about the parameter through the effective beam splitter. Whenever the derivative modes $\frac{if'_k}{w_k}$ have nonvanishing overlap with some initially unpopulated modes $f_k$, part of the information about the parameter will end up in vacuum modes whose fluctuations limit the measurement precision. To see this explicitly, note that we may rewrite Eq.~(\ref{eq:QFI}) as $F_Q[\hat{\rho}(\theta)]|_{\theta=0}=F_Q[\hat{\rho},\hat{H}_I]+\langle \hat{O}\rangle_{\hat{\rho}}$, where
\begin{align}\label{eq:vacobservable}
    \hat{O}=4\sum_{kl\in I}(f'_k|\Pi_{\mathrm{vac}}|f'_{l})\hat{a}_k^{\dagger}\hat{a}_l,
\end{align}
and $\Pi_{\mathrm{vac}}=\sum_{j\notin I}|f_j)(f_j|$ is the mode projector on all the vacuum modes.

\section{Quantum-enhanced strategies} 
\textbf{Mode design.} We are in the position to make a simple but important observation about the origin of quantum sensitivity scaling enhancements in mode parameter estimations: Any improvement of the measurement precision must have its origin in the unitary QFI $F_Q[\hat{\rho},\hat{H}_I]$, which is the only term that actually depends on the quantum fluctuations of the state. However, this term vanishes when the Hamiltonian~(\ref{eq:HI}) is zero, and in this case the SQL cannot be overcome. This can be avoided if there exist $k,l\in I$ such that $i(f_j|f'_k)\neq 0$.

In other words, a necessary condition for quantum-enhanced mode parameter estimation can be formulated as follows: For at least one initially populated mode $f_k$, a mode with nonzero overlap with the derivative mode $\frac{if_k'}{w_k}$ must also be populated. There are generally two ways to achieve this: (1) the mode $f_k$ may already be nonorthogonal to its own derivative mode $\frac{if_k'}{w_k}$, or (2) one may populate additional auxiliary modes that are proportional to $\frac{if_k'}{w_k}$.

In situation (1), quantum enhancements are possible even if $f_k$ is the only populated mode. The most general single-mode scenario is discussed in detail in the Supplemental document. We demonstrate that single-mode approaches are sufficient to achieve quantum-enhanced estimation of a mode parameter only if this parameter is encoded in the phase of the mode, which applies to the estimation of frequency and time, as well as to orbital angular momentum of Laguerre-Gauss modes~\cite{AmbrosioNatCommun2013}.

In practical situations, it may not always be possible to manipulate the shape of the modes of interest as they are usually determined by the problem at hand. Nevertheless, even when $f_k$ is orthogonal to its derivative mode $\frac{if_k'}{w_k}$, we may achieve quantum enhancements by following approach (2). To this end, we employ a multimode setting by incorporating suitable auxiliary modes with nonvanishing overlap with $\frac{if_k'}{w_k}$. The unitary QFI $F_Q[\hat{\rho},\hat{H}_I]$ increases as the overlap between the populated modes $f_k$ and their derivatives $\frac{if_k'}{w_k}$ grows. An extreme situation is found when all of the derivative modes $\frac{if_k'}{w_k}$ can be expanded using only the initially populated modes $f_k$. In this case, of which a Mach-Zehnder interferometer is an important instance (see Supplemental document), the second line in Eq.~(\ref{eq:QFI}) vanishes and all the information about the precision limit is contained in the unitary QFI.

In summary, the population of suitable modes allows us to establish the necessary condition for achieving quantum-enhanced measurement precision. However, this condition is not sufficient: These modes must also be populated with suitable nonclassical states in order to overcome the fluctuations of the vacuum.

\textbf{State design.} Quantum states that lead to a sensitivity improvement beyond the SQL can be identified for any nonzero effective Hamiltonian using standard methods from quantum metrology by maximization of the QFI over a set of quantum states under appropriate constraints~\cite{GiovannettiNATPHOT2011,PezzeRMP2018}. The choice of suitable nonclassical states depends on the limitations of the experimental setup at hand. Maximal quantum enhancements that achieve the Heisenberg limit typically require large and fragile superposition states that are hard to prepare, such as NOON states. Nevertheless, other classes of more accessible states are also able to achieve useful and scalable quantum enhancements under realistic conditions. For instance, we demonstrate in the Supplemental document that a strongly populated coherent state in a mode that is orthogonal to its own derivative can always be complemented by a squeezed state in a suitably designed auxiliary mode in order to improve the measurement precision of any mode parameter with simple homodyne measurements.

\section{Applications}
Given any mode parameter estimation task, a suitable measurement strategy is identified in two steps. First, a study of the effective Hamiltonian identifies a set of modes whose nonvanishing population establishes the necessary condition to achieve quantum-enhanced precision. Second, the precision limit for any quantum state, pure or mixed, Gaussian or non-Gaussian, prepared in those modes can be determined by virtue of the QFI. In the following, we apply our formalism to transverse spatial modes and superresolution imaging, focusing on the design of suitable modes. Additional examples are given in the Supplemental document.

\textbf{Transverse spatial modes.} Whenever the mode of interest is orthogonal to its own derivative, quantum enhanced precision can only be achieved by populating an additional auxiliary mode. This is the case for the measurement of transverse displacements of a beam described by Hermite-Gauss modes $f_n[\theta](x,y)=\mathrm{HG}_{nm}(x+\theta,y)$ with $m$ fixed (see Supplemental document for details), which is a fundamental task in optics known as beam positioning. Populating only the single mode $\mathrm{HG}_{nm}$ thus leads to the precision $F_Q[\hat{\rho}(\theta)]|_{\theta=0}=4\frac{(2n+1)}{w^2}\langle\hat{N}\rangle_{\hat{\rho}}$ that scales linearly with $N$ and depends on the beam waist $w$. A classical enhancement of the sensitivity is offered by modes of higher order $n$. By complementing with suitable auxiliary modes, the quadratic scaling and the potential for quantum enhancements, can be recovered. These auxiliary modes correspond to the derivatives $w\frac{\partial}{\partial x}\mathrm{HG}_{nm}=\sqrt{n}\mathrm{HG}_{n-1,m}-\sqrt{n+1}\mathrm{HG}_{n+1,m}$, and can again be expressed in terms of Hermite-Gauss modes. For arbitrary multimode quantum states that occupy a basis of Hermite-Gauss modes, we obtain the effective Hamiltonian~(\ref{eq:H})
\begin{align}
    \hat{H}=\frac{i}{w}\sum_{n}\sqrt{n+1}(\hat{a}^{\dagger}_n\hat{a}_{n+1}-\hat{a}^{\dagger}_{n+1}\hat{a}_n),    
\end{align}
demonstrating that information about the spatial displacement $\theta$ will leak from each mode $\mathrm{HG}_{nm}$ into the neighboring modes $\mathrm{HG}_{n-1,m}$ and $\mathrm{HG}_{n+1,m}$. Preparing these modes in nonclassical states with reduced quantum fluctuations therefore allows us to obtain a quantum-enhanced measurement precision. 
A similar analysis in the Supplemental document identifies strategies for quantum-enhanced estimation of the beam waist parameter $w$.


\textbf{Superresolution imaging.} The resolution of two point sources with a diffraction-limited imaging system is a mode parameter estimation problem of fundamental relevance for astronomy and microscopy~\cite{TsangPRX2016,TsangREVIEW2020}. The ultimate quantum limit was derived for thermal sources~\cite{TsangPRL2016,LupoPRL2016} but a general upper sensitivity bound reveals that even strongly nonclassical states cannot yield a quantum scaling enhancement~\cite{LupoPRL2016}. Following Refs.~\cite{TsangPRX2016,LupoPRL2016}, we describe the problem in an orthogonal mode basis containing (anti-)symmetric combinations $f_{\pm}$ of the local source modes. This orthogonalization procedure leads to parameter-dependent populations that are not described by the mode transformation due to losses~\cite{LupoPRL2016}. We thus amend our general expression~(\ref{eq:QFI}) by adding the classical Fisher information of the populations, $F_c=\sum_{n}(p'_n)^2/p_n$~\cite{BraunsteinPRL1994,ParisINTJQI2009}.

Whenever the phase of the point-spread function (PSF) of the imaging system~\cite{GoodmanBOOK} is independent of the transverse coordinate, we have $(f'_\mp|f'_\pm)=(f_\mp|f'_\pm)=(f_\pm|f'_\pm)=0$. Since the derivative modes are orthogonal to the populated modes, the vanishing of the Hamiltonian~(\ref{eq:H}) implies SQL scaling even if the source modes could be prepared in arbitrary nonclassical states~\cite{LupoPRL2016}. The remaining terms in Eq.~(\ref{eq:QFI}) read $F_Q[\hat{\rho}(\theta)]_{\theta=0}=F_c+4(f'_+|f'_+)\langle\hat{N}_+\rangle+4(f'_-|f'_-)\langle\hat{N}_-\rangle$ and produce exactly the expression that was derived in Ref.~\cite{LupoPRL2016} for number-diagonal states, demonstrating its validity for arbitrary states whose eigenstates are independent of the source separation (see Supplemental document).

Our general theory for quantum mode parameter estimation allows us to discuss possibilities for achieving beyond-SQL quantum enhancements in superresolution imaging, assuming that some control is available over the state of the sources as is the case, e.g., in certain microscopy settings or in the time-frequency domain. First, if the phase of the PSF is nonconstant, suitably constructed (anti-)symmetric modes will no longer be orthogonal to their derivatives. While for superresolution of spatial modes in the paraxial regime the assumption of a constant phase is well justified~\cite{GoodmanBOOK,LupoPRL2016}, nonconstant phases emerge naturally in the time-frequency domain, which has been studied experimentally~\cite{DonohuePRL2018}. For example, we show in the Supplemental document that a linear phase $\psi(x)=u(x)e^{-ikx}$ adds to the sensitivity the term $F_Q[\hat{\rho},\hat{H}]$ with the Hamiltonian~(\ref{eq:H}) $\hat{H}=k(\hat{N}_++\hat{N}_-)/2$. Because $F_Q[\hat{\rho},\hat{H}]=0$ whenever $\hat{\rho}$ and $\hat{H}$ commute, the sensitivity of number-diagonal states, in particular thermal states, remains unaffected by this additional term, reflecting their inability to overcome the SQL. However, suitable nonclassical emitters are able to exploit this term to achieve nonlinear sensitivity scalings with the number of photons $N$. Second, as we have seen in our general discussions as well as in previous examples, another possibility to achieve quantum scaling enhancements even if the mode shape cannot be modified consists in populating the derivative modes. These strategies open up interesting avenues for quantum-enhanced superresolution.

\section{Conclusions}
Any mode parameter estimation problem can be modeled by a suitable effective bilinear Hamiltonian that contains information about the shape of the modes. This reveals the general quantum limits for high-precision measurements of mode parameter, without requiring assumptions about specific states, modes, measurement observables or estimators. Our general result predicts precisely how the shape of modes influences the quantum limits and ultimately determines whether or not quantum enhancements beyond the standard quantum limit are possible. We find that, generally, such quantum enhancements can be achieved by a suitable design of the modes on which the probe state is being prepared. These results reveal strategies to optimize precision measurements of mode properties in quantum optical settings with light and atoms, including in spectroscopy and imaging.


\textbf{Acknowledgments}
C.F. thanks J. Lundeen and N. Boroumand for stimulating discussions, and the University of Ottawa for a visiting research position. This work received funding from the European Union’s Horizon 2020 Research and Innovation Programme under Grant Agreement No. 899587. This work was funded by French ANR under the COSMIC project (ANR-19-ASTR-0020-01). This work was funded by MCIN/AEI/10.13039/501100011033 and the European Union “NextGenerationEU” PRTR fund [RYC2021-031094-I]. This work has been founded by the Ministry of Economic Affairs and Digital Transformation of the Spanish Government through the QUANTUM ENIA project call - QUANTUM SPAIN project, by the European Union through the Recovery, Transformation and Resilience Plan - NextGenerationEU within the framework of the Digital Spain 2026 Agenda, and by the CSIC Interdisciplinary Thematic Platform (PTI+) on Quantum Technologies (PTI-QTEP+).

\textbf{Disclosures} The authors declare no conflicts of interest.

\textbf{Data availability} No data were generated or analyzed in the presented research.

\clearpage

\onecolumngrid

\begin{center}\large{\textbf{Supplemental Document}}\end{center}
\renewcommand{\theequation}{S.\arabic{equation}}
\setcounter{equation}{0}

\section*{Supplement 1: Describing mode transformations by an effective beam-splitter evolution}\label{app:H}
We start by expressing the shifted modes in terms of the original, unshifted basis, i.e.,
\begin{align}\label{eq:fkdev}
f_{k}[\theta]=\sum_j(f_j|f_{k}[\theta])f_j.
\end{align}
The same expansion~(\ref{eq:fkdev}) applies to the mode creation operators~\cite{FabreRMP2020}. Hence, if $\hat{a}_k^{\dagger}[\theta]$ and $\hat{a}_k^{\dagger}=\hat{a}_k^{\dagger}[0]$ create an excitation in the mode $f_{k}[\theta]$ and $f_{k}$, respectively, we obtain
\begin{align}\label{eq:adag}
\hat{a}_k^{\dagger}[\theta]=\sum_j(f_j|f_{k}[\theta])\hat{a}_j^{\dagger}.
\end{align}
Subsequent applications of $\hat{a}_k^{\dagger}[\theta]$ to the vacuum creates Fock states in the perturbed modes, which in turn can be used to represent arbitrary quantum states $|\psi\rangle_{\theta}$, parametrized by $\theta$. Here, we assume the coefficients of the state in such a representation to be $\theta$-independent, i.e., changes of $\theta$ exclusively affect the modes. 

We are interested in the evolution of quantum states due to changes in the modes that they occupy, while the coefficients of the state remain constant. An arbitrary pure multimode state can be expressed in terms of a basis of multimode Fock states
\begin{align}
|\psi\rangle_{\theta}&=\sum_{n_1=0}^{\infty}\sum_{n_2=0}^{\infty}\cdots c_{n_1,n_2,\dots}|n_1\rangle_{1,\theta} |n_2\rangle_{2,\theta}\cdots,
\end{align}
where $|n_k\rangle_{k,\theta}$ is an $n_k$-photon Fock state in the mode $f_{k}[\theta]$. We thus consider the evolution of the state $|\psi_{\theta}\rangle$ that is due to changes in these modes at constant coefficients $c_{n_1,n_2,\dots}$. We obtain
\begin{align}
    &\quad\frac{\partial}{\partial \theta}|n_1\rangle_{1,\theta} |n_2\rangle_{2,\theta}\cdots\notag\\&=\frac{1}{\sqrt{n_1!n_2!\cdots}}\frac{\partial}{\partial \theta}(\hat{a}^{\dagger}_1[\theta])^{n_1}(\hat{a}^{\dagger}_2[\theta])^{n_2}\cdots|0\rangle\notag\\
    &=\frac{1}{\sqrt{n_1!n_2!\cdots}}\left(\sum_{k}\left(\frac{\partial}{\partial \theta}(\hat{a}^{\dagger}_k[\theta])^{n_k}\right)\prod_{\substack{l\\l\neq k}}(\hat{a}^{\dagger}_l[\theta])^{n_l}\right)|0\rangle\notag\\
    &=\frac{1}{\sqrt{n_1!n_2!\cdots}}\left(\sum_{k}\left(n_k(\hat{a}_k^{\dagger}[\theta])^{n_k-1}\left(\frac{\partial}{\partial \theta}\hat{a}_k^{\dagger}[\theta]\right)\right)\prod_{\substack{l\\l\neq k}}(\hat{a}^{\dagger}_l[\theta])^{n_l}\right)|0\rangle\notag\\
    &=\left(\sum_{k}n_k\left(\frac{\partial}{\partial \theta}\hat{a}_k^{\dagger}[\theta]\right)\sqrt{\frac{(n_k-1)!}{n_k!}}\frac{1}{\sqrt{n_k}}\hat{a}_k[\theta]\right)|n_1\rangle_{1,\theta} |n_2\rangle_{2,\theta}\cdots\notag\\
    &=\left(\sum_{k}\left(\frac{\partial}{\partial \theta}\hat{a}_k^{\dagger}[\theta]\right)\hat{a}_k[\theta]\right)|n_1\rangle_{1,\theta} |n_2\rangle_{2,\theta}\cdots,
\end{align}
where we used that $\hat{a}_k[\theta]|n_k\rangle_{k,\theta}=\sqrt{n_k}|n_k-1\rangle_{k,\theta}$. We thus find that
\begin{align}\label{eq:dpsiSupp}
    \frac{\partial}{\partial \theta}|\psi\rangle_{\theta}=\sum_{k}\left(\frac{\partial}{\partial \theta}\hat{a}_k^{\dagger}[\theta]\right)\hat{a}_k[\theta]|\psi\rangle_{\theta},
\end{align}
where
\begin{align}\label{eq:dadag}
    \frac{\partial}{\partial \theta}\hat{a}_k^{\dagger}[\theta]=\sum_j(f_j|f'_{k}[\theta])\hat{a}_j^{\dagger},
\end{align}
and the prime denotes the derivative with respect to $\theta$, i.e., $f'_{k}[\theta]=\frac{\partial}{\partial \theta}f_{k}[\theta]$. 
We can always absorb a shift of the value of $\theta$ into the definition of the initial mode basis $f_k$, such that without loss of generality, we focus on deviations from the value $\theta=0$. Denoting $|\psi\rangle=|\psi\rangle_{\theta=0}$, we obtain from Eq.~(\ref{eq:dpsiSupp})
\begin{align}\label{eq:effevolSUPP}
    \frac{\partial}{\partial \theta}|\psi\rangle=-i\hat{H}|\psi\rangle,
\end{align}
where we introduced the effective beam-splitter Hamiltonian
\begin{align}\label{eq:Hmeth}
    \hat{H}&=\left.i\sum_{k}\left(\frac{\partial}{\partial \theta}\hat{a}_k^{\dagger}[\theta]\right)\hat{a}_k[\theta]\right|_{\theta=0}\notag\\
    &=i\sum_{ij}\sum_{k}(f_i|f'_{k})\underbrace{(f_{k}|f_j)}_{\delta_{kj}}\hat{a}_i^{\dagger}\hat{a}_j\notag\\
    &=i\sum_{ij}(f_i|f'_j)\hat{a}_i^{\dagger}\hat{a}_j\notag\\
    &=\sum_{k}w_k\hat{d}_k^{\dagger}\hat{a}_k.
\end{align}
In those last steps we used Eqs.~(\ref{eq:adag}) and~(\ref{eq:dadag}), and we introduced the operator
\begin{align}\label{eq:derivativemode}
d^{\dagger}_k=\frac{i}{w_k}\left.\frac{\partial}{\partial \theta}\hat{a}_k^{\dagger}[\theta]\right|_{\theta=0}=\frac{i}{w_k}\sum_{j}(f_j|f'_{k})\hat{a}_j^{\dagger},
\end{align}
which creates a particle in the derivative mode $\frac{if'_{k}}{w_k}$, with $f'_{k}=\frac{\partial}{\partial \theta}f_{k}[\theta]|_{\theta=0}$ and the normalization constant $w_k=\sqrt{(f'_k|f'_k)}$. 

To verify that $\hat{H}$ is indeed Hermitian, recall that the $f_k[\theta]$ form a basis and thus satisfy $(f_k[\theta]|f_l[\theta])=\delta_{kl}$. Expanding the left-hand side up to first order at $\theta=0$ yields
\begin{align}
    (f_{k}[\theta]|f_{l}[\theta])&=\delta_{kl}+\theta\left[(f'_k|f_l)+(f_k|f'_l)\right]+\mathcal{O}(\theta^2).
\end{align}
We thus obtain the condition
\begin{align}
    (f'_k|f_l)+(f_k|f'_l)&=0,\label{eq:ucond1}
\end{align}
implying that
\begin{align}
    \hat{H}^{\dagger}&=-i\sum_{ij}(f'_i|f_j)\hat{a}_i^{\dagger}\hat{a}_j=\hat{H}.
\end{align}

\section*{Supplement 2: Quantum Fisher information for general mode transformations}
\label{app:QFI}
We now consider a quantum state that depends on the parameter $\theta$ via its eigenstates. As $\theta$ is varied, these eigenstates retain their coefficients, but the modes they occupy vary as described above:
\begin{align}\label{eq:rho}
    \hat{\rho}(\theta)=\sum_{n}p_n|\psi_n\rangle_{\theta}\langle \psi_n|_{\theta}.
\end{align}
The quantum Fisher information for estimations of the parameter $\theta$ is given by~\cite{BraunsteinPRL1994,ParisINTJQI2009,STothJPA2014,SPezzeSmerziReview}:
\begin{align}
    F_Q[\hat{\rho}(\theta)]=2\sum_{\substack{n,m\\p_n+p_m>0}}\frac{(p_n-p_m)^2}{p_n+p_m}|_{\theta}\langle \partial_{\theta}\psi_n|\psi_m\rangle_{\theta}|^2,
\end{align}
where $|\partial_{\theta}\psi_n\rangle_{\theta}=\frac{\partial}{\partial\theta}|\psi_n\rangle_{\theta}$. Focussing on the sensitivity of the state at $\theta=0$, we obtain~\cite{SKnyshPRA2011,STothJPA2014}
\begin{align}\label{eq:QFIH_Supp}
    F_Q[\hat{\rho}(\theta)]|_{\theta=0}&=2\sum_{\substack{n,m\\p_n+p_m>0}}\frac{(p_n-p_m)^2}{p_n+p_m}|\langle \psi_n|\hat{H}|\psi_m\rangle|^2\notag\\
    &=4\langle\hat{H}^2\rangle_{\hat{\rho}}-8\sum_{\substack{n,m\\p_n+p_m>0}}\frac{p_np_m}{p_n+p_m}|\langle \psi_n|\hat{H}|\psi_m\rangle|^2,
\end{align}
where we made use of Eq.~(\ref{eq:effevolSUPP}) and we denote $|\psi_n\rangle=|\psi_n\rangle_{\theta=0}$.

Notice that $\hat{H}$ contains not only modes that are initially populated, but also those that are in the vacuum. To distinguish between these modes, we denote by $I$ the set of modes that are initially populated by the eigenstates $|\psi_n\rangle$ of~(\ref{eq:rho}), i.e., $\hat{a}_k|\psi_n\rangle=0$ for all $k\neq I$ and for all $n$. We define [recall the definition of $\hat{d}_k$ in Eq.~(\ref{eq:derivativemode})]
\begin{align}\label{eq:Hrhovac}
\hat{H}=\underbrace{\sum_{k\in I}w_k\hat{d}_k^{\dagger}\hat{a}_k}_{\hat{H}_{\hat{\rho}}}+\underbrace{\sum_{k\not\in I}w_k\hat{d}_k^{\dagger}\hat{a}_k}_{\hat{H}_{\mathrm{vac}}}.
\end{align}
From $\hat{H}_{\mathrm{vac}}|\psi_n\rangle=0$ for all $n$, we now obtain for the first term in~(\ref{eq:QFIH_Supp}) that
\begin{align}\label{eq:H2}
\langle\hat{H}^2\rangle_{\hat{\rho}}&=\langle\hat{H}_{\hat{\rho}}^2+\hat{H}_{\hat{\rho}}\hat{H}_{\mathrm{vac}}+\hat{H}_{\mathrm{vac}}\hat{H}_{\hat{\rho}}+\hat{H}_{\mathrm{vac}}^2\rangle_{\hat{\rho}}\notag\\
&=\langle\hat{H}_{\hat{\rho}}^2\rangle_{\hat{\rho}}+\langle\hat{H}_{\mathrm{vac}}\hat{H}_{\hat{\rho}}\rangle_{\hat{\rho}}.
\end{align}
Using
\begin{align}
[\hat{a}_k,w_l\hat{d}_l^{\dagger}]=i\sum_{j}(f_j|f'_{l})[\hat{a}_k,\hat{a}_j^{\dagger}]=i(f_k|f'_{l}),
\end{align}
we obtain
\begin{align}\label{eq:Hvacterm}
\langle\hat{H}_{\mathrm{vac}}\hat{H}_{\hat{\rho}}\rangle_{\hat{\rho}}&=\sum_{k\not\in I}\sum_{l\in I} w_k w_l\langle\hat{d}_k^{\dagger}\hat{a}_k\hat{d}_l^{\dagger}\hat{a}_l\rangle_{\hat{\rho}}\notag\\
&=\sum_{k\not\in I}\sum_{l\in I}w_k w_l\langle \hat{d}_k^{\dagger}\hat{d}_l^{\dagger}\hat{a}_k\hat{a}_l\rangle_{\hat{\rho}}+i\sum_{k\not\in I}\sum_{l\in I}w_k(f_k|f'_{l})\langle \hat{d}_k^{\dagger}\hat{a}_l\rangle_{\hat{\rho}}\notag\\
&=-\sum_{k\not\in I}\sum_{l\in I}\sum_j(f_k|f'_{l})(f_j|f'_k)\langle \hat{a}_j^{\dagger}\hat{a}_l\rangle_{\hat{\rho}}\notag\\
&=\sum_{jl\in I}\sum_{k\not\in I}(f'_j|f_k)(f_k|f'_{l})\langle \hat{a}_j^{\dagger}\hat{a}_l\rangle_{\hat{\rho}}\notag\\
&=\sum_{jl\in I}\left[(f'_j|f'_{l})-\sum_{k\in I}(f'_j|f_k)(f_k|f'_{l})\right]\langle \hat{a}_j^{\dagger}\hat{a}_l\rangle_{\hat{\rho}},
\end{align}
and we used~(\ref{eq:ucond1}) and the completeness of the basis $f_k$.
Finally, note that $\langle \hat{H}_{\hat{\rho}}^2\rangle=\langle\hat{H}_{I}^2\rangle$ 
and $\langle \hat{H}_{\hat{\rho}}\rangle=\langle \hat{H}_{I}\rangle$ where
\begin{align}\label{eq:HI_Supp}
\hat{H}_I=i\sum_{jk\in I}(f_j|f'_k)\hat{a}_j^{\dagger}\hat{a}_k.
\end{align}
Inserting Eqs.~(\ref{eq:Hrhovac}), (\ref{eq:H2}) and~(\ref{eq:Hvacterm}) into Eq.~(\ref{eq:QFIH_Supp}) yields the final result, Eq.~(\ref{eq:QFI}).

A further generalization is obtained by considering that the eigenvalues $p_k(\theta)$ of the state~(\ref{eq:rho}) may also depend on the parameter $\theta$. This leads to an additional term in the quantum Fisher information that describes the classical Fisher information of the populations~\cite{BraunsteinPRL1994,ParisINTJQI2009,SPezzeSmerziReview}, and we obtain
\begin{align}\label{eq:QFIg}
    F_Q[\hat{\rho}(\theta)]|_{\theta=0}&=\sum_n\frac{1}{p_n}\left(\frac{\partial p_n}{\partial \theta} \right)^2+F_Q[\hat{\rho},\hat{H}_I]\\
    &\quad +4\sum_{kl\in I}\left[(f'_k|f'_{l})-\sum_{j\in I}(f'_k|f_j)(f_j|f'_{l})\right]\langle \hat{a}_k^{\dagger}\hat{a}_l\rangle_{\hat{\rho}},\notag
\end{align}
where $p_k=p_k(0)$ and $\frac{\partial}{\partial \theta} p_k=\frac{\partial}{\partial \theta} p_k(\theta)|_{\theta=0}$.

The result~(\ref{eq:QFIg}) determines the quantum limit on the estimation of an arbitrary mode parameter as a function of the shape of the populated modes. Being valid for arbitrary quantum state, our explicit expression for the quantum Cram\'er-Rao bound can be used to identify optimal quantum states and measurements for arbitrary mode parameter. In the following we illustrate the applicability of this result by providing the quantum limits on the estimation of a selection of mode parameter of particular interest.

\section*{Supplement 3: Examples}
This section provides additional details on the examples discussed in the main manuscript as well as further illustrative case studies of the general mode parameter estimation framework.

\subsection*{Single-mode case} 
Let us focus on the case where only a single mode $f_{\theta}(x)=A_{\theta}(x)e^{-i\phi_{\theta}(x)}$ is initially occupied, with all remaining modes in the vacuum. This scenario contains the common situation of experiments that use a single-mode laser for precision measurements, e.g., of frequency or spatial and temporal positioning. We assume that amplitude $A_{\theta}$ and phase $\phi_{\theta}$ are real, differentiable functions of $\theta$. From Eqs.~(\ref{eq:HI_Supp}) and~(\ref{eq:QFIg}), we obtain $\hat{H}_I=i(f|f')\hat{N}$ and
\begin{align}\label{eq:singlemodesens}
    F_Q[\hat{\rho}(\theta)]|_{\theta=0}=|(f|f')|^2F_Q[\hat{\rho},\hat{N}]+4[(f'|f')-|(f|f')|^2]\langle\hat{N}\rangle_{\hat{\rho}}, 
\end{align}
where $\hat{N}=\hat{a}^{\dagger}\hat{a}$ is the number operator of the mode $f=f_{\theta=0}$. The relevant mode integrals read
\begin{align}\label{eq:singlemodemodes}
    (f|f')&=-i\int dxA(x)^2\phi'(x),\notag\\
    (f'|f')&=\int dx\left[A'(x)^2+A(x)^2\phi'(x)^2\right],
\end{align}
where we used the short notation $F(x)=F_{\theta}(x)|_{\theta=0}$ and $F'(x)=\frac{\partial}{\partial \theta}F_{\theta}(x)|_{\theta=0}$ for $F=A,\phi$.

Since the term $F_Q[\hat{\rho},\hat{N}]$ in Eq.~(\ref{eq:singlemodesens}) scales quadratically in $\hat{N}$, it can be improved with squeezed vacuum and it is maximized by the strongly nonclassical superposition states $(|0\rangle+|N\rangle)/\sqrt{2}$. This is, however, only possible when the effective Hamiltonian~(\ref{eq:HI_Supp}), $\hat{H}_I=i(f|f')\hat{N}$, is nonzero. From Eq.~(\ref{eq:singlemodemodes}) we observe that this can be achieved only when the phase $\phi_{\theta}(x)$ is a nonconstant function of $\theta$ at $\theta=0$. Any mode with a parameter-independent phase is orthogonal to its own derivative and therefore the sensitivity~(\ref{eq:singlemodesens}) reduces to
\begin{align}\label{eq:singlemodesensconst}
F_Q[\hat{\rho}(\theta)]|_{\theta=0}=4(f'|f')\langle\hat{N}\rangle_{\hat{\rho}}.
\end{align}
For a shifted Gaussian beam $f_{\theta}(x)=f(x+\theta)$ with $f(x)=A(x)=(2/\pi w^2)^{1/4}e^{-x^2/w^2}$, we obtain that $(f'|f')=w^{-2}$ is determined by the variance, i.e., the beam waist when $x$ is a transverse spatial coordinate or the frequency spread when $f$ describes the spectrum. The expression~(\ref{eq:singlemodesensconst}) holds for arbitrary states and shows that the parameter sensitivity is always limited to a linear scaling in $N=\langle\hat{N}\rangle_{\hat{\rho}}$, independently of the initial state or measurement scheme that is used. 

This is different when information about the parameter of interest is contained in the phase. For a translation of a linear phase $f_{\theta}(x)=A_{\theta}(x)e^{-ik(x+\theta)}$, we have $\phi_{\theta}(x)=k(x+\theta)$ and $\hat{H}_I=k\hat{N}$, independently of the amplitude $A_{\theta}$. If additionally $A_{\theta}$ is independent of $\theta$, the derivative mode has zero overlap with the vacuum modes and we obtain that $F_Q[\rho(\theta)]|_{\theta=0}=k^2F_Q[\hat{\rho},\hat{N}]$. This situation applies, e.g., to the estimation of orbital angular momentum with Laguerre-Gauss modes, which has been explored experimentally in Ref.~\cite{AmbrosioNatCommun2013}. In addition to potential quantum gains, higher-excited modes with larger $k$ lead to a classical sensitivity improvement by a factor of $k^2$.



\subsection*{Transverse spatial modes}
For the estimation of transverse spatial beam properties, we consider a basis of Hermite-Gauss modes, defined as
\begin{align}\label{eq:HG} 
\mathrm{HG}_{nm}(x,y;z,w,k)&=\sqrt{\frac{2}{\pi2^{n+m}n!m!}}\frac{1}{w(z)}H_n\left(\frac{\sqrt{2}x}{w(z)}\right)H_m\left(\frac{\sqrt{2}y}{w(z)}\right)\notag\\&\quad\times e^{-\frac{x^2+y^2}{w(z)^2}}e^{-ik\frac{x^2+y^2}{2R(z)}}e^{-ikz}e^{i(n+m+1)\psi(z)},
\end{align}
with
\begin{align}
    \begin{array}{lrl}
w(z)=w\sqrt{1+\left(\frac{z}{z_R}\right)^2},&\qquad\psi(z)&=\mathrm{arctan}\left(\frac{z}{z_R}\right),\\
R(z)=z+\frac{z_R^2}{z},&z_R&=\frac{w^2k}{2},
    \end{array}
\end{align}
with wave vector $k$, beam waist $w$, and $H_n(x)=(-1)^ne^{x^2}\frac{d^n}{dx^n}e^{-x^2}$ are the Hermite polynomials. We discuss the estimation of displacements in the transverse coordinates $x$ or $y$ and of the beam waist $w$. The corresponding effective Hamiltonian~(\ref{eq:Hmeth}) depends on the derivative of these modes with respect to the parameter of interest.

\textbf{Transverse beam displacements}
We consider first transverse displacements of a beam. For simplicity, we consider the displacement axis to be the $x$ axis and we focus on the excited modes in the $x$ direction. We define a mode basis $f_{n}[\theta](x,y)=\mathrm{HG}_{n,m_0}(x+\theta,y;z,w,k)$ for some arbitrary, fixed value of $m_0$. We use basic properties of the Hermite polynomials $H_n(x)$, namely 
the recurrence relation
\begin{align}
    H_{n+1}(x)=2xH_n(x)-H_{n}'(x),
\end{align}
and
\begin{align}
    H_n'(x)=2nH_{n-1}(x),
\end{align}
to express the derivative of Hermite-Gauss modes as
\begin{align}\label{eq:HGderivative}
    w\frac{\partial}{\partial x}\mathrm{HG}_{nm}=\sqrt{n}\mathrm{HG}_{n-1,m}-\sqrt{n+1}\mathrm{HG}_{n+1,w}.
\end{align}
The orthonormality of the Hermite-Gauss modes implies that
\begin{align}
    w(f_n|f'_m)=\sqrt{m}\delta_{n,m-1}-\sqrt{m+1}\delta_{n,m+1},
\end{align}
and
\begin{align}
    w^2(f'_n|f'_m)&=(2n+1)\delta_{n,m}-\sqrt{n(n-1)}\delta_{n,m+2}\notag\\&\quad-\sqrt{(n+1)(n+2)}\delta_{n,m-2}.
\end{align}
The effective beam-splitter Hamiltonian, Eq.~(\ref{eq:H}), thus reads
\begin{align}\label{eq:HGHamiltonian}
    \hat{H}=\frac{i}{w}\sum_{n}\sqrt{n+1}(\hat{a}^{\dagger}_n\hat{a}_{n+1}-\hat{a}^{\dagger}_{n+1}\hat{a}_n),
\end{align}
and the sum extends over all values of $n\geq 0$. The existence of quantum enhancements, i.e., quadratic terms, now again depends on the populations in the modes $f_n(x,y)$.


To take full advantage of the Hamiltonian~(\ref{eq:HGHamiltonian}), we need nonvanishing population in pairs of modes with indices $n$ and $n+1$. The simplest scenario that allows for quantum enhancements is given when $n_{\max}=1$, i.e., when besides the fundamental mode only the first excited mode is populated. For the fundamental Gaussian mode $f_0(x,y)=\mathrm{HG}_{00}(x,y)=\sqrt{2/\pi w^2}e^{-(x^2+y^2)/w^2}$, we obtain $f'_0(x,y)=-\mathrm{HG}_{10}(x,y)/w$, which is orthogonal to $f_0$. This scenario has been studied theoretically~\cite{FabreOPTLETT2000,TrepsPRA2005,PinelPRA2012} and realized experimentally with squeezed vacuum~\cite{STrepsPRL2002,TrepsSCIENCE2003}. Population of the derivative mode $f_1=-wf'_0(x,y)$ leads to a sensitivity as large as
\begin{align}\label{eq:disp}
    w^2F_Q[\hat{\rho}(\theta)]|_{\theta=0}=F_Q[\hat{\rho},i(\hat{a}_0^{\dagger}\hat{a}_1-\hat{a}_1^{\dagger}\hat{a}_0)]+8\langle\hat{a}^{\dagger}_1\hat{a}_1\rangle_{\hat{\rho}}.
\end{align}

\textbf{Beam waist}
For the estimation of the beam waist $w$, we define $f_{nm}[\theta](x,y)=\mathrm{HG}_{nm}(x,y;z,w+\theta,k)$. A calculation similar to the one above reveals that
\begin{align}
    &\quad 2w\frac{\partial}{\partial w}\mathrm{HG}_{nm}=-\sqrt{n(n-1)}\mathrm{HG}_{n-2,m}-\sqrt{m(m-1)}\mathrm{HG}_{n,m-2}\notag\\&+\sqrt{(n+2)(n+1)}\mathrm{HG}_{n+2,m}+\sqrt{(m+2)(m+1)}\mathrm{HG}_{n,m+2}.\notag
\end{align}
The evolution of the state under variations of the beam waist around $w$, i.e., changes of $\theta$, where the beam waist is given by $w+\theta$, is described by the effective Hamiltonian
\begin{align}
    \hat{H}
    &=\frac{i}{2w}\sum_{n,m}\left(\hat{c}_{n,m}^{\dagger}\hat{a}_{n,m}-\hat{a}_{n,m}^{\dagger}\hat{c}_{n,m}\right),\label{eq:Hbeamwaist}
\end{align}
where $\hat{a}^{\dagger}_{n,m}$ creates a photon in the mode $\mathrm{HG}_{nm}$ and
\begin{align}
    \hat{c}_{n,m}=\sqrt{(n+2)(n+1)}\hat{a}_{n+2,m}+\sqrt{(m+2)(m+1)}\hat{a}_{n,m+2}.\notag
\end{align}

Clearly, populating only a single $\mathrm{HG}$ mode, again, leads to an SQL-limited estimation precision, since all modes are orthogonal to their derivatives. If only the mode $\mathrm{HG}_{nm}$ for fixed choices of $n$ and $m$ is populated, we obtain the sensitivity limit
\begin{align}
F_Q[\hat{\rho}(\theta)]|_{\theta=0}=\frac{2(n^2+m^2+n+m+2)}{w^2}\langle \hat{N}\rangle_{\hat{\rho}}.
\end{align}
which is linear in $N$ and, for the fundamental mode, $n=m=0$, coincides with the limit for displacement sensing with the same mode. In contrast, quadratic terms may emerge when modes proportional to $\hat{c}_{n,m}$ are populated. For example, a quantum state in the fundamental mode $f_{00}$ can be complemented by population of the normalized derivative mode $wf'_{00}=(\mathrm{HG}_{20}+\mathrm{HG}_{02})/\sqrt{2}$, which leads to a sensitivity of
\begin{align}\label{eq:waist}
    w^2F_Q[\hat{\rho}(\theta)]|_{\theta=0}=F_Q[\hat{\rho},i(\hat{c}_{00}^{\dagger}\hat{a}_{00}-\hat{a}_{00}^{\dagger}\hat{c}_{00})]+16\langle\hat{c}^{\dagger}_{00}\hat{c}_{00}\rangle_{\hat{\rho}},
\end{align}
where $\hat{c}^{\dagger}_{00}$ creates a photon in the mode $wf'_{00}$. The second term in~(\ref{eq:waist}) again reflects losses to the vacuum, since we cannot express the second mode's derivative mode, $f''_{00}$, in terms of the two initially populated modes $f_{00}$ and $wf'_{00}$.

\subsection*{Mean field mode}
In this simple but general scenario, we illustrate how to optimally design both a classical mode and a quantum state that will maximize the precision of a mode parameter measurement in a realistic setting. We consider the experimentally common situation in which some mode $f_0$ is prepared in a strongly populated coherent state $|\alpha_0\rangle$ with $|\alpha_0|^2=N\gg 1$. We further assume that $(f_0|f'_0)=0$, which applies to relevant precision measurements, such as interferometers or estimations of transverse spatial displacements or the beam waist. We address the question: how can the sensitivity be optimized under these constraints by making optimal use of the remaining modes, i.e., what kind of state should we prepare and in which modes?

We devide the set $I$ of initially populated modes into $I=\{0\}\cup I_+$, where we assume that the mode $k=0$ is prepared in the state $|\alpha_0\rangle$, leading to the total initial state $\hat{\rho}=|\alpha_0\rangle\langle \alpha_0|\otimes\hat{\rho}_+$, where $\hat{\rho}_+$ describes the state on the modes in $I_+$. In order to determine $F_Q[\hat{\rho}(\theta)]|_{\theta=0}$ from Eq.~(\ref{eq:QFI}), we first consider $F_Q[\hat{\rho},\hat{H}_I]$. According to Eq.~(\ref{eq:QFIH_Supp}), we need to determine the terms $|\langle\Psi_n|\hat{H}_I|\Psi_m\rangle|^2$ and $\langle \hat{H}_I^2\rangle_{\hat{\rho}}$. Considering the first term, we note that the eigenstates of $\hat{\rho}$ are given by $|\Psi_n\rangle=|\alpha_0\rangle\otimes|\varphi_n\rangle$, where $|\varphi_n\rangle$ are the eigenvectors of $\hat{\rho}_+$. We thus obtain
\begin{align}
    |\langle\Psi_n|\hat{H}_I|\Psi_m\rangle|^2&=N|\langle\varphi_n|e^{-i\phi}\hat{b}+e^{i\phi}\hat{b}^{\dagger}|\varphi_m\rangle|^2+\mathcal{O}(\sqrt{N}),
\end{align}
where we have introduced
\begin{align}
    \hat{b}&=i\sum_{k\in I_+}(f_0|f'_k)\hat{a}_k,\\
    \alpha_0&=\sqrt{N}e^{i\phi}.
\end{align}
Moreover, we obtain that
\begin{align}
    \langle \hat{H}^2_I\rangle_{\hat{\rho}}&=N\langle (e^{-i\phi}\hat{b}+e^{i\phi}\hat{b}^{\dagger})^2\rangle_{\hat{\rho}_+}+\mathcal{O}(\sqrt{N}).
\end{align}
Finally, in the second term in Eq.~(\ref{eq:QFI}), only terms where $k=l=0$ will contribute at the leading order in $N$ and thus, we finally obtain
\begin{align}\label{eq:QFImf0}
    \frac{1}{N}F_Q[\hat{\rho}(\theta)]|_{\theta=0}&=F_Q[\hat{\rho}_+,e^{-i\phi}\hat{b}+e^{i\phi}\hat{b}^{\dagger}]+4\left[(f'_0|f'_0)-\sum_{k\in I_+}(f'_0|f_k)(f_k|f'_0)\right]+\mathcal{O}\left(\frac{1}{\sqrt{N}}\right).
\end{align}
We note that $[\hat{b},\hat{b}^{\dagger}]=\sum_{k\in I_+}(f'_0|f_k)(f_k|f'_0)$ and introduce the physical, i.e., normalized mode
\begin{align}
    \hat{c}=\frac{1}{\sqrt{\sum_{k\in I_+}(f'_0|f_k)(f_k|f'_0)}}\hat{b},
\end{align}
which, by construction, satisfies $[\hat{c},\hat{c}^{\dagger}]=1$. We express the quadrature operator of the normalized mode $\hat{c}$ as 
\begin{align}
    \hat{q}_{\phi}&=e^{-i\phi}\hat{c}+e^{i\phi}\hat{c}^{\dagger}\notag\\
    &=\frac{1}{\sqrt{\sum_{k\in I_+}(f'_0|f_k)(f_k|f'_0)}}(e^{-i\phi}\hat{b}+e^{i\phi}\hat{b}^{\dagger}),
\end{align}
and we obtain from Eq.~(\ref{eq:QFImf0}):
\begin{align}\label{eq:QFImf}
    \frac{1}{N}F_Q[\hat{\rho}(\theta)]|_{\theta=0}&=\sum_{k\in I_+}|(f'_0|f_k)|^2F_Q[\hat{\rho}_+,\hat{q}_{\phi}]+4\left[(f'_0|f'_0)-\sum_{k\in I_+}|(f'_0|f_k)|^2\right]+\mathcal{O}\left(\frac{1}{\sqrt{N}}\right).
\end{align}
This expression reveals optimal choices for both the mode and the corresponding quantum state. To see this, let us consider two extreme scenarios. 

First notice that the second term is bounded from above by $4(f'_0|f'_0)$, since the sum is positive. This upper bound is reached when all elements in the sum are zero, which is the case when $I_+$ is the empty set, i.e., besides the mean field mode, no other mode is populated, or when all the other populated modes $f_k$ with $k\in I_+$ are orthogonal to $f'_0$. In this case, the sensitivity is given by 
\begin{align}\label{eq:FQmfvac}
    F_Q[\hat{\rho}(\theta)]|_{\theta=0}=4N(f'_0|f'_0).    
\end{align}
This expression is independent of $\hat{\rho}_+$ and, hence, this scenario leaves no further possibilities for optimizations over the quantum state. This is simply due to the fact that whatever modes are populated besides the mean field mode $f_0$, they carry no information about the parameter and thus their quantum state is irrelevant for the task at hand.

The opposite extreme is described by the scenario where $f'_0$ can be perfectly reconstructed as linear combinations of the populated modes $f_k$ with $k\in I_+$. In this case, we have $(f'_0|f'_0)=\sum_{k\in I_+}|(f'_0|f_k)|^2$. A simple possibility to achieve this is to populate the mode $f'_0/\sqrt{(f'_0|f'_0)}$. We obtain
\begin{align}
    F_Q[\hat{\rho}(\theta)]|_{\theta=0}&=N(f'_0|f'_0)F_Q[\hat{\rho}_+,\hat{q}_{\phi}].
\end{align}
This expression depends on the state of the modes in $I_+$ and it is maximized by a pure state because of the convexity of $F_Q$ in the state. For a pure state $\hat{\rho}_+=\hat{\psi}_+\equiv|\psi_+\rangle\langle\psi_+|$, we obtain
\begin{align}
    F_Q[\hat{\rho}(\theta)]|_{\theta=0}&=4N(f'_0|f'_0)(\Delta\hat{q}_{\phi})^2_{\hat{\psi}_{+}}.
\end{align}
We thus see that pure states with a variance $(\Delta\hat{q}_{\phi})^2_{\hat{\psi}_{+}}$ above the value of $1$ outperform the sensitivity of Eq.~(\ref{eq:FQmfvac}). Indeed, the vacuum state satisfies $(\Delta\hat{q}_{\phi})^2_{|0\rangle\langle 0|}=1$, confirming once more the result~(\ref{eq:FQmfvac}), as expected, for the case when only the mean field mode is populated. Improvements beyond this bound are possible by using squeezed vacuum states with sub-SQL quantum noise along the conjugate quadrature $\hat{q}_{\phi+\pi}$---a standard technique in quantum optics. Moreover, among all states with the same average photon number, this strategy maximizes the quantum enhancement and is therefore optimal.

\subsection*{Mach-Zehnder interferometer}
A closed system of two modes with no information loss to the vacuum is a Mach-Zehnder interferometer. Even though this case is well understood~\cite{CavesPRD1981,SPezzePRL2000,GiovannettiNATPHOT2011}, it is instructive to treat it in our framework of mode parameter estimation. Assume that we estimate the phase $\theta$ from the mode
\begin{align}
    f_1=\frac{1}{\sqrt{2}}(g_1e^{-i\theta/2}+g_2e^{i\theta/2}),
\end{align}    
where $g_1,g_2$ are two orthonormal modes describing the two arms of the interferometer. The derivative mode
\begin{align}
f'_1=-\frac{i}{2\sqrt{2}}(g_1e^{-i\theta/2}-g_2e^{i\theta/2})
\end{align}
is orthogonal to $f_1$, implying a sensitivity limited to the SQL whenever one interferometer input mode is in the vacuum. By populating also
\begin{align}
f_2=\frac{1}{\sqrt{2}}(g_1e^{-i\theta/2}-g_2e^{i\theta/2}),
\end{align}
we obtain $f'_1=(-i/2)f_2$ and $f'_2=(-i/2)f_1$. Hence, the vacuum term, Eq.~(\ref{eq:vacobservable}), disappears and we obtain that
\begin{align}
    \hat{H}_I=\frac{1}{2}(\hat{a}_2^{\dagger}\hat{a}_1+{a}_1^{\dagger}\hat{a}_2).    
\end{align}
The sensitivity $F_Q[\hat{\rho}(\theta)]|_{\theta=0}=F_Q[\hat{\rho},\hat{H}_I]$ scales quadratically in $\hat{H}_I$ and can therefore be improved beyond the SQL using appropriate nonclassical states, such as squeezed states~\cite{CavesPRD1981,SPezzePRL2000,GiovannettiNATPHOT2011}.



\subsection*{Timing resolution with chirped pulses}
Consider a temporal mode described by the chirped pulse
\begin{align}\label{eq:chirped}
    f(t;a,b,\omega_0)=\left(\frac{2 a}{\pi}\right)^{1/4} e^{-i \omega_0 t - (a + i b)t^2},
\end{align}
with $a,\omega_0>0$, $b\in\mathbb{R}$. Such pulses are used in radar systems to estimate the distance of an object by measuring the time of arrival of the reflected pulse~\cite{SKlauderBELL1960}. We thus consider estimations of a time delay $f[\theta](t;a,b,\omega_0)=f(t+\theta;a,b,\omega_0)$. Assuming that only the mode~(\ref{eq:chirped}) is initially populated, with all other modes in the vacuum, the sensitivity limit is described by Eq.~(\ref{eq:singlemodesens}). We obtain
\begin{align}
    i(f|f')&=\omega_0,\notag\\
    (f'|f')&=a+\frac{b^2}{a}+\omega_0^2,
\end{align}
leading to the quantum limit
\begin{align}
    F_Q[\hat{\rho}(\theta)]|_{\theta=0}=\omega_0^2F_Q[\hat{\rho},\hat{N}]+4\left(a+\frac{b^2}{a}\right)\langle\hat{N}\rangle_{\hat{\rho}}.
\end{align}

\section*{Supplement 4: Superresolution imaging}
This section contains the details on the application of our framework to the scenario of superresolution imaging.

We formulate the problem of superresolution imaging~\cite{TsangPRX2016,TsangPRL2016,LupoPRL2016} in the framework of mode parameter estimation with two populated modes. Consider two incoherent point sources at positions $\pm s/2$. The emitted light is collected by a diffraction-limited imaging system~\cite{GoodmanBOOK} with point-spread function $\psi(x)$. In general the PSF
\begin{align}
    \psi(x)=e^{i\varphi(x)}u(x),
\end{align}
is described by real functions for the amplitude $u(x)$ and phase $\varphi(x)$, with normalization $\int dx u^2(x)=1$.

To circumvent the problem of nonorthogonality of the distributions generated by the two sources in the image plane, $\psi(x\pm s/2)$, we introduce the orthonormal symmetric and anti-symmetric modes
\begin{align}\label{eq:fpm}
f_{\pm}(x)=\frac{\psi(x+s/2)\pm \frac{\delta}{|\delta|}\psi(x-s/2)}{\sqrt{2(1\pm|\delta|)}},
\end{align}
which we can extend to a full spatial mode basis. Here
\begin{align}\label{eq:deltageneral}
    \delta&=\int dx\psi^*(x-s/2)\psi(x+s/2)\notag\\
    &=\int dx e^{i\varphi(x+s/2)-i\varphi(x-s/2)}u(x-s/2)u(x+s/2).
\end{align}
For a PSF with constant phase, i.e., when $\varphi(x)$ is independent of $x$, this coincides with the ansatz by Lupo and Pirandola~\cite{LupoPRL2016}. 

We are interested in estimating a variation of $s$, i.e., here our parameter $\theta$ of interest describes a displacement $s\to s+\theta$ and will be estimated in the vicinity of $\theta=0$. Assuming that only the two sources are initially populated, we obtain that only the modes $\hat{a}_{\pm}$ are populated in the image plane and $I=\{+,-\}$ in Eq.~(\ref{eq:QFIg}).

We now focus on the special case where the phase dependence is linear, i.e., when
\begin{align}
    \psi(x)=e^{-ikx}u(x).
\end{align}
In this case, we obtain from Eq.~(\ref{eq:deltageneral})
\begin{align}
    \delta&=e^{-iks}\underbrace{\int dx u(x-s/2)u(x+s/2)}_{|\delta|},
\end{align}
and the (anti-)symmetric modes~(\ref{eq:fpm}) are given by
\begin{align}\label{eq:fpmlin}
    f_{\pm}(x)=e^{-ik(x+s/2)}\frac{u(x+s/2)\pm u(x-s/2)}{\sqrt{2(1\pm|\delta|)}}.
    \end{align}
To identify the quantum limit, we need to identify the mode overlap integrals. The derivative modes read
\begin{align}\label{eq:SRdetectionmodes}
\frac{\partial f_{\pm}(x)}{\partial s}=\left(-i\frac{k}{2}\mp\frac{\gamma}{2(1\pm|\delta|)}\right)f_{\pm}(x)+\frac{\frac{d\psi(x+s/2)}{dx}\mp \frac{d\psi(x-s/2)}{dx}}{2\sqrt{2(1\pm|\delta|)}},
\end{align}
where
\begin{align}
    \gamma=\frac{\partial |\delta|}{\partial s}.
\end{align}
Using that $u(x)$ is normalized and vanishes at $\pm\infty$, we obtain
\begin{align}\label{eq:SRoverlaps}
    (f_\pm|f'_\pm)&=-i\frac{k}{2},\notag\\
    (f_\mp|f'_\pm)&=0,\notag\\
    (f'_\pm|f'_\pm)&=\frac{k^2}{4}+\frac{1}{4(1\pm|\delta|)}\left((\Delta p)^2\mp\beta\right)-\frac{\gamma^2}{4(1\pm|\delta|)^2},\notag\\
    (f'_\mp|f'_\pm)&=0,
\end{align}
where we introduced
\begin{align}
(\Delta p)^2&=\int dx\left|\frac{d \psi(x)}{d x}\right|^2=\int dx\left(\frac{d u(x)}{dx}\right)^2,\\
\beta&=\int dx \frac{d u(x-s/2)}{dx}\frac{d u(x+s/2)}{dx}.
\end{align}

The quantum Fisher information for sub-wavelength imaging of two sources in arbitrary quantum states is given according to Eq.~(\ref{eq:QFIg}) as
\begin{align}\label{eq:QFISR}
    F_Q[\hat{\rho}(\theta)]|_{\theta=0}
    &=\sum_{n,m}\frac{1}{p_{n,m}}\left(\frac{\partial p_{n,m}}{\partial \theta} \right)^2+F_Q[\hat{\rho},\frac{k}{2}(\hat{a}_{+}^{\dagger}\hat{a}_{+}+\hat{a}_{-}^{\dagger}\hat{a}_{-})]\notag\\&\quad+4\left[(f'_{+}|f'_{+})-|(f'_{+}|f_+)|^2\right]\langle \hat{a}_{+}^{\dagger}\hat{a}_{+}\rangle_{\hat{\rho}}\notag\\&\quad+4\left[(f'_{-}|f'_{-})-|(f'_{-}|f_-)|^2\right]\langle \hat{a}_{-}^{\dagger}\hat{a}_{-}\rangle_{\hat{\rho}},
\end{align}
where we used the orthogonality described in Eqs.~(\ref{eq:SRoverlaps}). The nonvanishing coefficients are independent of $k$ and read
\begin{align}\label{eq:coeffsfpm}
    4\left[(f'_{\pm}|f'_{\pm})-|(f'_{\pm}|f_{\pm})|^2\right]=\frac{1}{1\pm|\delta|}\left[(\Delta p)^2\mp\beta-\frac{\gamma^2}{1\pm|\delta|}\right].
\end{align}
This generalizes previous results on the sensitivity limit in quantum imaging that were available for a PSF with constant phase, i.e., $k=0$, assuming thermal states~\cite{TsangPRL2016,LupoPRL2016}, or more generally, states that are diagonal in the Fock basis~\cite{LupoPRL2016}. Interestingly, for nonzero $k$, the unitary QFI generated by the nonvanishing Hamiltonian 
\begin{align}
    \hat{H}_I=\frac{k}{2}(\hat{a}_{+}^{\dagger}\hat{a}_{+}+\hat{a}_{-}^{\dagger}\hat{a}_{-}),
\end{align}
can be exploited with common nonclassical strategies to suppress the quantum measurement noise below the standard quantum limit.

Let us finally show how to recover the results of Refs.~\cite{TsangPRL2016,LupoPRL2016} from our general result~(\ref{eq:QFISR}) as a special case. Thermal states with $\eta N$ photons in the nonorthogonal modes correspond to thermal states in the symmetric/anti-symmetric modes with average photon numbers~\cite{LupoPRL2016}
\begin{align}\label{eq:npm}
    \langle \hat{a}_{\pm}^{\dagger}\hat{a}_{\pm}\rangle_{\hat{\rho}}=N_{\pm}=\eta(1\pm |\delta|)N.
\end{align}
The thermal phonon populations
\begin{align}
    p_{n,m}=\frac{1}{N_{+}+1}\frac{1}{N_{-}+1}\left(\frac{N_{+}}{N_{+}+1}\right)^n\left(\frac{N_{-}}{N_{-}+1}\right)^m
\end{align}
depend on the parameter $\theta$ and yield the classical Fisher information~\cite{LupoPRL2016}
\begin{align}\label{eq:Fc}
    F_c=\sum_{n,m}\frac{1}{p_{n,m}}\left(\frac{\partial p_{n,m}}{\partial \theta} \right)^2&=2\eta N\left[\frac{\gamma^2}{2(1+\delta)(1+(1+\delta)\eta N))}+\frac{\gamma^2}{2(1-\delta)(1+(1-\delta)\eta N))}\right].
\end{align}
Inserting Eqs.~(\ref{eq:coeffsfpm}), (\ref{eq:npm}), and~(\ref{eq:Fc}) into Eq.~(\ref{eq:QFISR}), we obtain for $k=0$
\begin{align}
    F_Q[\hat{\rho}(\theta)]|_{\theta=0}&=2\eta  N  \left((\Delta p)^2 - \frac{\eta N(1+\eta N)\gamma^2}{(1 + \eta N)^2 -\eta^2 N^2 \delta^2}\right).
\end{align}
This result coincides with the one given in Refs.~\cite{TsangPRL2016,LupoPRL2016}.

\end{document}